\documentclass[12pt,preprintnumbers,aps,amssymb,nofootinbib,amsmath]{revtex4}
\usepackage{epsfig,epsf}
\usepackage{graphicx}
\usepackage{epstopdf}
\newcommand{\beq}{\begin{equation}}
\newcommand{\eeq}{\end{equation}}
\def\eq#1{{(\ref{#1})}}
\def\fig#1{{Fig.~\ref{#1}}}

\newcommand{\as}{\alpha_s}

\newcommand{\im}{\mathrm{Im}}

\newtheorem{theorem}{Lecture~\#\!\!}
\newcommand{\bet}{\begin{theorem}}
\newcommand{\eet}{\end{theorem}}
\newcommand{\be}{\begin{eqnarray}}
\newcommand{\ee}{\end{eqnarray}}

\newcommand{\lb}{\left(}
\newcommand{\rb}{\right)}

\begin{document} 

\preprint{BNL-NT-06/8, TAUP 2821/06}

\title{Multi-particle Production and Thermalization\\ in High-Energy QCD}

\author{Dmitri Kharzeev$^a$ ,  Eugene Levin$^{b}$ and Kirill Tuchin 
$^{c,d}$}
 
\affiliation{
a) Department of Physics, Brookhaven National Laboratory,\\
Upton, New York 11973-5000, USA\\
b) HEP Department, School of Physics,\\
Raymond and Beverly Sackler Faculty of Exact Science,\\
Tel Aviv University, Tel Aviv 69978, Israel\\
c) Department of Physics and Astronomy,\\
Iowa State University, Iowa, 50011, USA\\
d) RIKEN BNL Research Center,\\
Upton, New York 11973-5000, USA\\
}

\date{\today}
\pacs{}

\begin{abstract} 
We argue that multi--particle production in high energy hadron and nuclear collisions can be considered as proceeding through the production of gluons in the background classical field. In this approach we derive the gluon spectrum immediately after the collision and find that at high energies  it is parametrically enhanced by $\ln(1/x)$ with respect to the quasi--classical result ($x$ is the Bjorken variable). We show that  the produced gluon spectrum becomes thermal (in three dimensions) with an effective temperature determined by the saturation momentum $Q_s$, $T= c\ Q_s/2\pi$ during the time $\sim 1/T$; we estimate $c=\sqrt{2\pi}/2 \simeq 1.2$. Although this result by itself does not imply that the gluon spectrum will remain thermal at later times, it has an interesting applications to heavy ion collisions. In particular, we discuss the possibility of Bose--Einstein condensation of the produced gluon pairs and estimate the viscosity of the produced gluon system.

\end{abstract}

\maketitle

\section{Introduction}
Recently, it was suggested that a fast thermalization in heavy-ion collisions can occur through the gluon radiation off rapidly decelerating nuclei  \cite{KT}.  In that paper two of us have pointed out that a pulse of strong chromo-electric field produces Schwinger--like \cite{Schwinger:1951nm} radiation with a thermal spectrum.  We also discussed an analogy between the Schwinger mechanism and the Hawking--Unruh radiation and its application to heavy-ion collisions (see also \cite{Kerman:1985tj,Kluger:1991ib,Bialas:1999zg}).  The macroscopic approach of \cite{KT}  led to a number of intriguing but qualitative results. In the present paper we would like to reconcile the macroscopic approach of \cite{KT} with the microscopic one based on the QCD parton model.

\begin{figure}[ht]
    \begin{center}
        \includegraphics[width=0.90\textwidth]{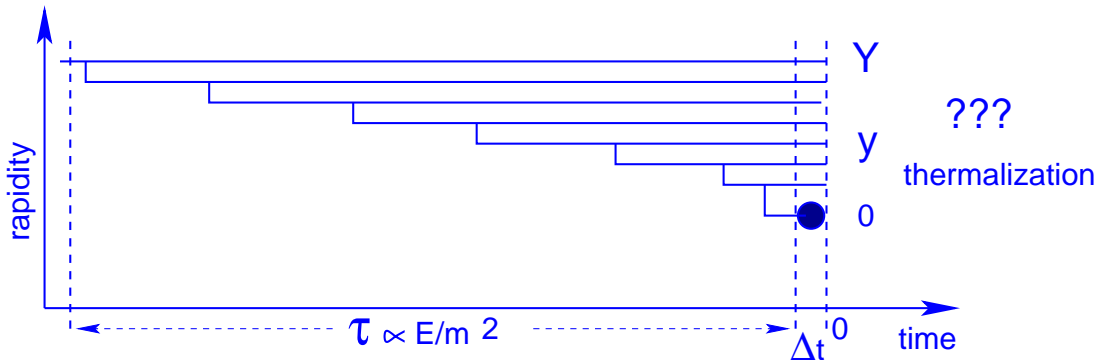}
\end{center}
\caption{The time structure of the parton cascade for a fast particle (nucleus) in the target rest frame. Note, that ordering of the high energy gluons in rapidity $y$ is equivalent to their ordering in the space-time rapidity $\eta=(1/2)\,\ln(x_+/x_-)$. In the target rest frame, the target's rapidity is $\eta=0$, while that of the fast projectile is $\eta=Y>0$. The relation between the light-cone coordinates of the fastest gluon in the cascade is therefore $x_+=x_-e^{2Y}$, while for the slowest one $x_+\ge x_-$.     
}
\label{timecgc}
\end{figure}

The main goal of this paper is to give a picture of the thermalization stage of the process of multiparticle 
production in heavy ion collisions at high energy in the framework of  the color glass condensate (CGC) approach 
to high 
density QCD \cite{GLR,MUQI,MV}. The CGC approach is based on two principle ideas. The first one is the structure of the parton 
cascade at high energy which is shown in \fig{timecgc}. The main contribution to the high energy scattering is given by 
a parton fluctuation in which all partons are strongly ordered in time.  Let $\hat z$ be the beam direction in the rest frame of the target.     
The typical lifetime
of this fluctuation at high energy of the projectile $\varepsilon$ is large and is proportional to $\varepsilon/m^2$, where $m$ is the virtuality of the fluctuation. In terms of the light-cone variables $k^{\pm}\,= \varepsilon_i \pm k_{zi}$ the life-time of 
the $i$-th parton is of the order of  $t_i \equiv x_{+i}=\,1/k_{i}^{-}\,=\,k_{i}^{+}/k^2_{i\bot}$, where $k_{i\bot}$ is the 
transverse momentum of the $i$-th parton. Introducing the rapidity $y_i$ of the parton we can rewrite the lifetime as $t_i \,=\,(1/k_{i\bot})\cdot e^{y_i}$.  

The interaction with the target of the size $R$ destroys the coherence of the parton wave function of the 
projectile. The typical time, which is needed for this, is of the order of $\Delta t\sim R$ and is much smaller than the 
lifetime of all faster partons in the fluctuation:  $\Delta t \ll t_i$. Therefore, this interaction cannot change the momentum distribution of the fast parton in the projectile wave function. The influence of the target mostly manifests itself  in the loss of 
coherence for majority of the partons;  changes in momenta occur only for a few very slow (`wee') partons. The `wee' parton part of the wave function together with the interaction with the target could be factorized 
out while the energy dependence and distributions of the fast partons should not depend on the properties of the 
target. (In the following discussion we will assume that the interaction happens at time  $t=0$.)
Such picture follows from the parton model and is based on rather general properties of field theories (see e.\ g.\ 
ref.~\cite{PWF}); it has been proven in QCD for the BFKL emission \cite{BFKL}. 

The CGC approach adds a very essential new idea to the parton cascade picture. Since all partons with 
rapidity larger than $y$ (see \fig{timecgc})   live longer  than the parton with rapidity $y$, for a dense system 
such as a nucleus 
they can be considered as the source of the classical field that emits a gluon with rapidity $y$ \cite{MV}.
We are going to explore this idea to evaluate the parton wave function at the time $t =0$ (see \fig{timecgc}), or 
 to say better just after the interaction, when the coherence of the wave function has been destroyed (see section~\ref{sec:2}). 
Moreover, we will argue in Section~\ref{sec:2} that  the dominant source of 
parton production is the \emph{longitudinal} background field; we will also elucidate the origin of this field.

We then use the same background field approximation to follow the parton system at later times.
 In Section 3 we will argue that the produced parton spectrum assumes the three-dimensional thermal form (in a co-moving frame, of course) 
over the time $\sim 1/Q_s$, where $Q_s$ is the saturation scale which is a  
new dimensional parameter that characterizes the partonic wave function at $t= 0$ \cite{GLR,MUQI,MV}.
 We confirm the result of \cite{KT} that the effective  temperature is approximately $T = Q_s/(2\pi)$. 
 At later times, the partons will interact with each other and these interactions finally could 
create a thermalized system of partons in the true "thermodynamical" sense (for example, with temperature related to the density by equation of state), but the consideration of this late kinetic equilibration stage is beyond the scope of this paper. We would like to note only that a three-dimensional 
thermal shape of parton distributions should make a true kinetic equilibration easier.

In this paper we will use also two other key properties of a dense partonic system in QCD:

 The first one is the 
appearance of a new scale (saturation momentum $Q_s$) \cite{GLR,MUQI,MV}  which  characterizes the mean transverse momentum of 
partons in the parton cascade. This momentum is proportional to the density of partons (gluons) in the projectile  at  fixed rapidity,
namely, $Q^2_s \propto xG(x,Q_s)/\pi R^2$ where $xG$ is the number of gluons with fixed Bjorken $x=\exp(-y)$  and $R$ is the transverse size of the projectile. This scale increases with rapidity since $xG \propto (1/x)^\lambda$ 
in the region of low $x$. It means that  the smaller is the value of $Q_s$ the faster is the parton. Therefore, the 
parton with rapidity $y$ in \fig{timecgc} has a mean transverse momentum which is much larger than the transverse 
momentum of all partons moving faster than it; thus it can be considered as a probe for the system of  
fast partons, similar to the deep inelastic probe. This observation allows us to consider the production of a parton as a process of  emission 
by the frozen system of faster partons; averaging over the quantum numbers of incoming hadrons can be done after calculating the cross section. The typical 
configuration of the emitter is such that the transverse sizes are much larger (transverse momenta are much smaller) 
than the typical transverse sizes for the emitted parton (transverse momentum of emitted parton).

The second main idea behind the CGC approach is that the quantum emission in each stage of the process should give the same result as the emission by the classical field. This idea is the cornerstone of the Wilson  renormalization group approach 
in JIMWLK formalism (see Ref. \cite{JIMWLK}). 
\vskip0.3cm

This paper is organized as follows. In Sec.~\ref{sec:2} we explain the origin of the longitudinal fields in high energy hadron and nuclear collisions. We then consider the motion of a gluon in the external longitudinal  color field. The  imaginary part of the gluon propagator in an external field is related to the cross section for the inclusive  gluon production, Eq.\eq{FTOUR}.  We calculate the gluon propagator for an arbitrary external field in Sec.~\ref{sec:gp} using the WKB approximation. In Sec.~\ref{sec:evol} we use the derived formulae to calculate the imaginary part of the gluon propagator. Depending on the value of the adiabaticity parameter $\gamma$, see \eq{somedef}, we obtain the gluon spectrum at early times \eq{SOL9} and at later times \eq{TER6}. These are the main results of our paper. Eq.~\eq{SOL9} coincides with the McLerran-Venugopalan formula \cite{MV} for gluon emission by dense randomly distributed two-dimensional color charges.  The corresponding saturation scale is given by \eq{QS}.
Eq.~\eq{TER6} implies that at later times gluon distribution is thermal with the temperature determined by the saturation scale \eq{sattemp}.  Assuming the validity of $k_\bot$ factorization, in Sec.~\ref{sec:AA} we generalize our formalism to the case of heavy ion collisions. In Sec.~\ref{sec:therm} we consider multiple gluon pair production. Since the gluon spectrum at later times is thermal we apply well-known formalism of statistical physics to calculate the thermal properties of the produced gluon system. In particular, we observe the phenomenon of Bose-Einstein condensation which may solve the long-standing puzzle of multiple soft gluon production. 
 We discuss and summarize our results in 
Sec.~\ref{sec:discussion}.

 \section{High energy particle production by external fields}\label{sec:2}

\subsection{Transverse and longitudinal fields of the CGC}\label{sec:trlon}

The potential of a charge moving with constant velocity $v$ along the $z$-axis is given by a particular case of
 the Lienard-Weichert potential (see e.\ g.\  \cite{Mueller:1988xy})

\beq\label{pott1}
A_\pm\,=\,\frac{ 1\pm v}{\sqrt{2}} \frac{g/4\pi}{\sqrt{\frac{1}{2}[x_+(1-v)-x_-(1+v)]^2+(1-v^2)x_\bot^2}}
\eeq
\beq\label{pott2}
\vec A_\bot\,=\,0\,,
\eeq
where we introduced the light-cone potential $A^\mu=(A_+,A_-,\vec A_\bot)$ with $A_\pm=(A_0\pm A_z)/\sqrt{2}$.
If the particle is fast, then $v\rightarrow 1$ and the potential takes form
\beq\label{pott3}
A_+\,=\,\frac{\sqrt{2}g/4\pi}{\sqrt{2x_-^2+(1-v^2)x_\bot^2}}\,,
\eeq
\beq\label{pott4}
A_-\,=\,\vec A_\bot\,=\,0\,.
\eeq

The corresponding fields read
\beq\label{field2}
\vec E_\bot\,=\, \frac{g}{4\pi}\,\frac{(1-v^2)\vec x_\bot}{(2x_-^2+(1-v^2)x_\bot^2)^{3/2}}\,,
\eeq
\beq\label{field3}
\vec H\,=\,\vec v\,\times\,\vec E\,.
\eeq
Dirac equation in the background field \eq{pott3}-\eq{pott4} was solved in Ref.~\cite{volkov} with an assumption that the fast particle moves freely from $x_+=-\infty$ to $x_+=\infty$.  In this case 
the potentials \eq{pott3}-\eq{pott4} generate purely transverse mutually orthogonal electric and magnetic fields.  The action for such a plane wave background  vanishes. This implies that there is no pair production in a single monochromatic plane wave background \cite{Schwinger:1951nm}. 

The initial conditions in our case are different. As explained in the caption of \fig{timecgc}, for any gluon in the cascade it holds that $x_+\ge x_-$ or, equivalently, $z\ge 0$. In other words, in the target rest frame, all gluons move in the same positive $z$ direction. 
Therefore, the potential $A_+$ exists only in the positive half-plane $z\ge 0$. In other words we have to solve the pair production problem with the initial condition which explicitly depends on both lightcone coordinates $x_+$ and $x_-$. In Sec.~\ref{sec:EOM} we show 
that such an initial condition generates the longitudinal chromoelectric field $E_z$ in addition to the transverse fields mentioned above 
(see \eq{brb1} - \eq{brb3} and below).  The existence of longitudinal fields in the Color Glass Condensate has been pointed out previously 
in  Ref.~\cite{Kovner:1995ja}. In Ref.~\cite{Bialas:1986mt,Kerman:1985tj,Kluger:1991ib,Gatoff:1987uf} the pair production mechanism in heavy-ion collisions by non-perturbative fields has been discussed.

The longitudinal field $E_z$ is not only generated in a high energy collision, but it gives a leading contribution to the pair production amplitude as we are now going to demonstrate.  Consider  a system of fast  charges located 
at coordinates $\vec x_i$ randomly distributed in the transverse area of typical size $R_\bot$. Let us calculate a 
field created by all these charges at the point $x^\mu$. Assuming for simplicity  a continuous distribution of the charge,  
in the leading order in the coupling we have \beq\label{sum}
A_+(x)\,=\,\int \frac{d^3x}{4\pi}\frac{\rho(\vec x', t-|\vec x-\vec x'|)}{|\vec x-\vec x'|}\,=\,
\int \frac{d^2x'_\bot dx_-'}{4\pi}\frac{\sqrt{2}\,\rho(x_-',x_\bot',x_++x_-'-\frac{(x_\bot-x_\bot')^2}{2(x_--x_-')})}
{x_--x_-'+\frac{(x_\bot-x_\bot')^2}{2(x_--x_-')}}\,.
\eeq
Typical partons having rapidities  $y$ and $y'$ such that $y'>y$  have $x_-$'s satisfying $x_-\gg x_-' $. Also, the typical transverse size of  a parton decreases down the cascade as $x_\bot \sim 1/Q_s(y)$ since $Q_s(y)$ is an exponentially increasing function of $y$. Therefore the transverse sizes satisfy $x_\bot'\gg x_\bot$ which implies that the field $A(x)$ does not depend on the transverse size of the parton $x_\bot$:
\beq\label{sum2}
A_+(x)\,\approx \, \int \frac{d^2x'_\bot dx_-'}{4\pi}\frac{\sqrt{2}\,\rho(x_-',x_\bot',x_++x_-'-\frac{(x_\bot')^2}{2x_-})}{x_-+\frac{(x_\bot')^2}{2x_-}}\,.
\eeq
Eq.~\eq{sum2} implies that at high energies the transverse fields experienced by the partons are small compared to the longitudinal ones:  
\beq\label{sum2a}
|\vec E_\bot|\,=\,|\vec H_\bot|\,\ll\, E_z\,.
\eeq

We need to consider the result of \eq{sum2a} with  some caution since $E_\bot$ is still enhanced at 
very small values of $x_-$, see \eq{pott3}. However in the Lagrangian $\mathcal{L} = (E^2 - H^2)/2$ 
the transverse fields indeed give a very  small contribution proportional to $E^2_\bot - H^2_\bot 
\approx E_z^2\, Q^2_s(y')/Q^2_s(y)\ll E^2_z$. 

The pair production probability is proportional to the imaginary part of the effective Lagrangian evaluated 
by considering the quantum fluctuations in the background of the external color fields. Therefore, we expect 
that the pair production will be dominated by the longitudinal color fields; we will check this by an 
explicit calculation below.

\subsection{Particle production in the background field}\label{sec:pp}

\begin{figure}
    \begin{center}
       \includegraphics[width=13cm]{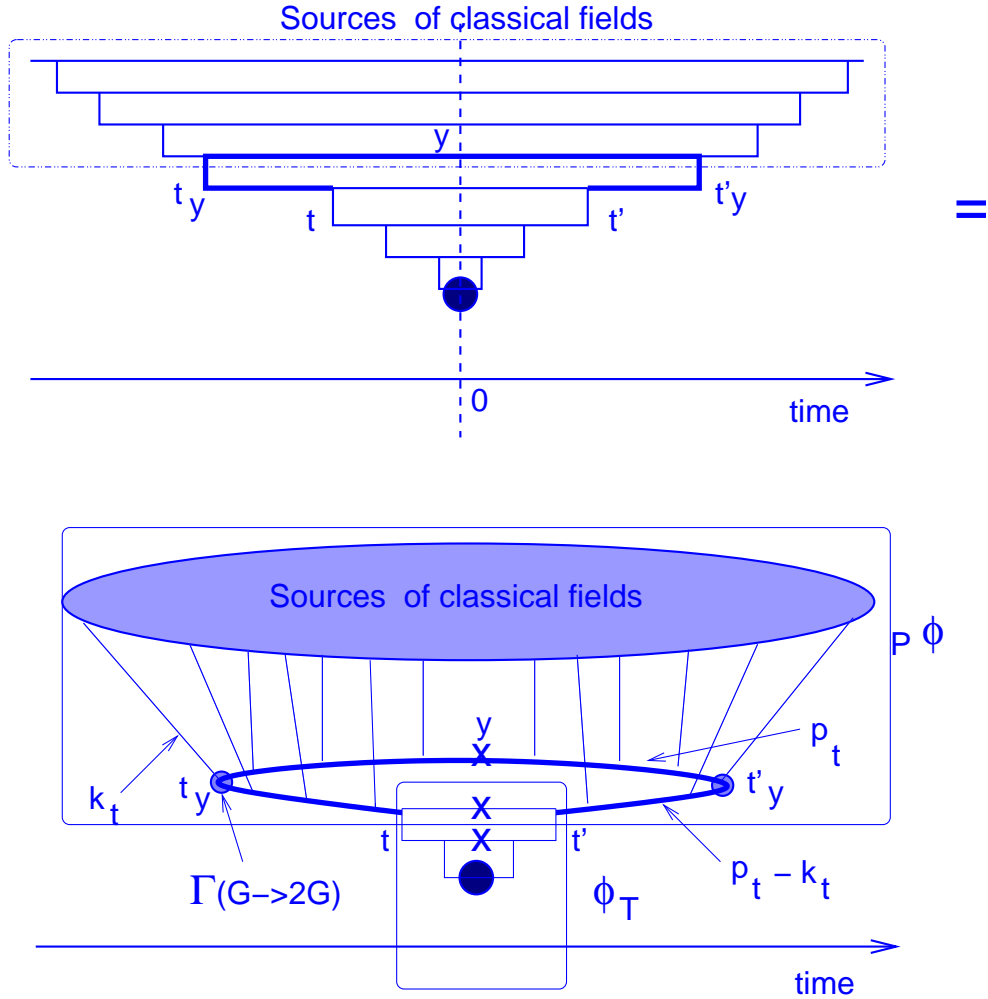}
 \end{center}
\caption{Inclusive gluon production at rapidity $y$ in the target rest frame. Crosses mark gluons which are on mass shell.  $\varphi_T$ and $\varphi_P$ denote the gluon densities for the target and the projectile, respectively.  The pair of gluons produced at rapidity $y$ at time $t_y$ in the amplitude and $t'_y$ in the complex conjugated one is shown by a bold line. 
}
\label{timeclsfield}
\end{figure}

Inclusive production of a gluon  with rapidity $y$ in a gluon cascade shown in \fig{timeclsfield} can be considered 
as a production of a gluon in a constant background field. Indeed,
for this gluon all other gluons with larger rapidities are effectively frozen and constitute a constant 
classical field $E_z$. Therefore, the splitting of a fast gluon into two gluons at rapidity $y$ at the time $t_y$ ($t'_y$ in the complex conjugated amplitude)  can be considered as a process of a gluon pair production by the field $\vec E$.  As shown in \fig{timeclsfield} both gluons propagate in the classical background field. Interaction with the target takes much shorter time than the gluon emission $t_y - t'_y \,\gg\,t - t'$. Therefore, the only dynamical role of the interaction with the target is to break the coherence of the nuclear wave function and to allow an inclusive measurement. This is the reason why 
we can present the inclusive cross section in a factorized form, namely,
$d \sigma/d y\,\propto \,\varphi_P(t_y - t'_y)\cdot \varphi_T(t - t')$, where $\varphi_P$ is the probability to find a 
gluon in 
the projectile. Calculating this probability we could neglect the fact that one gluon interacts with the target because of the short interaction time.  The unintegrated distribution $\varphi_T$ is thus the probability for a gluon to interact with the target.  Clearly, this simple factorization formula is just another representation for the well--known $k_t$-factorization 
formula which holds in high density QCD, at least for the interaction of a nucleus  with a virtual photon or hadron 
targets,  
\cite{KOTU} and which has  the form (see
Refs.\cite{GLR,LR1,LL,KR,GM,KM,BRAUN,K00} )
\beq \label{GENX}
\varepsilon \frac{d\sigma}{d^3 p} = \frac{4 \pi N_c}{ N_c^2 - 1}\frac {1}{ p_\bot^2}
 \int d k_\bot^2 \
\alpha_S \, \varphi_{P}(Y - y, k_\bot^2)\, \varphi_{T}(y,
(p-k)_\bot^2),
\eeq
In our approach it is convenient to write this formula in a different way
\beq \label{FTOUR}
\varepsilon \frac{d \sigma}{d^3 p} \,=\,\int \,d^2 \,k_\bot \Gamma^2(G\to 2G) \,\im D(Y - y,\vec{p}_\bot - \vec{k}_\bot)\, \im D(Y - y, \vec{k}_\bot)
\,\varphi_T(y,\vec{p}_\bot - \vec{k}_\bot)
 \eeq
where
\beq\label{verex}
\Gamma^2(G\to 2G)\,=\,\as\,\frac{4 \pi N_c }{ N_c^2 - 1}               
\eeq
and $\im D(y,k_t)$ is the imaginary part of the gluon propagator in the strong classical field. 

Let us consider a target with the transverse size $R $ much smaller than $1/Q_s(y)$ where $Q_s$ is the saturation momentum. For example,  consider the virtual photon target with virtuality $Q^2 \,\gg\,Q^2_s(y)$. In this case we can neglect the dependence on $k_t$ in $\varphi_T$ in 
\eq{FTOUR} and write 
\beq \label{FTOUR1}
 \frac{d \sigma}{d y d^2 p_\bot} \,=\,\varphi_{P}(Y - y, p_\bot)\,\varphi_{T}(y, p_\bot)
\eeq
with
\beq \label{PHIPR}
\varphi_{P}(Y - y, p_\bot)\,=\,\int \,d^2 \,k_\bot \Gamma^2(G\to 2G)\, \im D(Y-y,\vec{p}_\bot - \vec{k}_\bot)\, \im D(Y - y, \vec{p}_\bot)
\eeq

The dependence on $k_\bot$ is absorbed in the dependence of the classical fields on the transverse coordinate. In the first approximation we consider 
the classical fields being independent of the transverse coordinate. It means that the gluon propagator is proportional  to $\delta^{(2)}( \vec{k}_\bot)$ and \eq{PHIPR}
can be rewritten as
\beq \label{PHIPR1}
\varphi_{P}(Y - y, p_\bot)\,=\, \Gamma^2(G\to 2G)\, \im D(Y-y,\vec{p}_\bot )\, \im D(Y - y, \vec{p}_\bot)
\eeq
The factor $1/p^2_\bot$ is included in our definition of $D(Y - y, \vec{p}_\bot)$; it must be reproduced for the values of $p_\bot$ at which the perturbation theory is valid. 
\subsection{Equation of motion in the background field}\label{sec:EOM}

Now we can concentrate our efforts on the calculation of $\varphi_P$ which describes the production of the 
gluon pair in the strong and constant field. This problem has been investigated in detail both in QED and 
QCD (see review 
\cite{DUNNE} and references therein) and can be solved by using the background field method. Let us assume that gluon fields have the 
structure
\beq \label{GF}
G_\mu \,=\,A_\mu\,+\,W_\mu\,,
\eeq
where $A_\mu$ is a classical background field and $W_\mu$ is a quantum fluctuation.
The QCD Lagrangian can be expanded around this classical field and it has the following general structure \cite{THOOFT}
\beq \label{LAG}
{\cal L}[A + W]\,=\,{\cal L}[A]\,+\,\frac{\partial {\cal L}[A]}{\partial 
A_\mu}\,W_\mu\,+\,\frac{1}{2}\,\frac{\partial^2 {\cal L}[A]}{ 
\partial A_\mu\,\partial A_\nu}\,W_\mu \,W_\nu
\eeq
Since the second term is equal to zero due to equation of motion for the classical field, our Lagrangian has a 
quadratic form as far as the quantum field dependence is concerned. In the case of $SU(2)$ an explicit calculation (see Appendix~\ref{appendix}) leads to the equation of motion 
for the quantum field $W_\mu$:
\beq \label{EQM}
\lb - (\partial_\lambda \,-\,i g A_\lambda)^2\,\delta_{\mu \nu}\,+\,2i\,g\,F_{\mu \nu}[A] \rb \, W_\mu\,=\,0\,,
\eeq 
where we used the gauge condition $D_\mu A^\mu=0$. The field configuration discussed in Sec.~\ref{sec:trlon} satisfies this condition since $\partial_\mu A^\mu=\partial_- A_+(x_-)=0$ and $A_\mu^2=0$. 

The tensor $F_{\mu \nu}[A]$ for the longitudinal  electrical fields is
\beq \label{EQM1}
F_{\mu \nu}\,=\,\left(\begin{array} { c c c c }
0 & 0 & 0 & E_z \\
0 & 0 & 0 & 0  \\
0 & 0 & 0 & 0  \\
-E_z & 0 & 0 & 0 \\
\end{array}
\right)\,,
\eeq
where $E_z=-\partial_- A_+$.  The components of \eq{EQM} look as follows
\be
\mu = 0; & - (\partial_\lambda -i g A_\lambda)^2 \,W_0 + i g E_z W_3 & = 0\,, \label{0C}\\
\mu = 3; & - (\partial_\lambda -i g A_\lambda)^2 \,W_3 - i g E_z W_0 & = 0\,.   \label{3C}
\ee
Introducing $W_{\pm}=W_0 \pm i W_3$ we can rewrite \eq{0C} and \eq{3C} in the form
\beq \label{EQMF}
- \lb\,(\partial_\lambda - i g A_\lambda)^2 \pm 2g E_z \,\rb W_{\pm} \,=\,0\,.
\eeq

\subsection{Calculation of a gluon propagator in the background field}\label{sec:gp}

We now turn to solving the Eq.~\eq{EQMF}. Although $A_+$ is a function of only $x_-$ \eq{EQMF} cannot be solved by separation of variables since the initial condition depends on both $x_+$ and $x_-$ as has been discussed in Sec.~\ref{sec:trlon}. We can only separate the $x_\bot$ dependence. 
Thus, we are looking for the solution in the form 
$W_\sigma =e^{-i S-ip_\bot \cdot x_\bot}$, where $\sigma=\pm 1$. 
Working in the WKB approximation $|\partial_+ S \partial_-S|\gg |\partial_+\partial_- S|$ \cite{popov,MAPO}
we reduce \eq{EQMF} to 
\beq\label{sem}
-2\partial_+ S (\partial_- S -g A_+(x_-)) + p_\bot^2 + 2g\,\sigma E_z  =0\,, \quad x_+\ge x_-\,,
\eeq
where $\partial_+=\frac{\partial}{\partial x_-}$ and $\partial_-=\frac{\partial}{\partial x_+}$.
Eq.~\eq{EQMF} is a Hamilton-Jacobi equation for motion of a charged particle in the background field $A_+=A_+(x_-)$. The only difference from the classical mechanics is the appearance of the spin-dependent term in the right hand side of \eq{sem}. 

In the Hamilton-Jacobi formalism the action $S$ is considered along the true trajectories (satisfying Hamilton equations). It is a function of the coordinate $x$ of the final point of the trajectory. The action along the true trajectories can be found using the method of characteristics. This method was suggested for this class of problems in \cite{GLR,SCSOL} (for a mathematical review see e.\ g.\ \cite{KAMKE}). 

\subsubsection{Solution with $\sigma=0$}

In the spinless case $\sigma=0$, characteristics of Eq.~\eq{sem} are given by the solution of the following set of ordinary differential equations valid at $x_+\ge x_-$
\begin{eqnarray}
\frac{dx_-}{dt}&=& -2p_-\,,\label{vv1}\\
\frac{dx_+}{dt}&=& -2(p_+-g A_+(x_-))\,,\label{vv2}\\
\frac{dS}{dt}&=&-2 p_+p_--2p_-(p_+-g A_+)\,,\label{vv3}\\
\frac{dp_-}{dt}&=& 0\,,\label{vv4} \\
\frac{dp_+}{dt}&=& -2p_-g A_+'(x_-)\label{vv5}\,,
\end{eqnarray}
where $t$ is a parameter along the characteristics and we introduced the canonical momenta $p_\pm$ as 
\beq\label{ppdef}
p_-=-\partial_- S\,,\,\quad p_+=-\partial_+ S\,.
\eeq
Instead of one of the equations \eq{vv1} - \eq{vv4} we can use the following equation stemming from \eq{sem} and \eq{ppdef}
\beq\label{ms}
-2p_-(p_++eA_+(x_-))+p_\bot^2\,=\,0\,.
\eeq 
We will use \eq{ms} in place of \eq{vv4}.

We can use $x_-$ as a parameter along the characteristics  and rewrite  \eq{vv2}, \eq{vv3} and \eq{vv5} in the following way 

\begin{eqnarray}
\frac{dx_+}{dx_-}&=& \frac{p_++gA_+(x_-)}{p_-},\label{ssop1}\\
\frac{dS}{dx_-}&=& 2p_++gA_+(x_-)\label{ssop2}\,,\\
\frac{dp_+}{dx_-}&=& -gA_+'(x_-)\,.\label{ssop3}
\end{eqnarray}

Using Eq.~\eq{ms} the system \eq{ssop1}-\eq{ssop3} can be easily integrated with the following result

\begin{eqnarray}
p_+&=& -gA_+(x_-)\,+\, gA_+(x_+)\,+\,p_+^0\,,\label{brb1}\\
x_-&=& \frac{p_\bot^2}{2}\int\frac{dx_+}{(p_+^0+gA_+(x_+))^2} \,,\label{brb2}\\
S&=& -\int gA_+(x_-)dx_- \,+\, \int dx_+\frac{p_\bot^2}{p_+^0+gA_+(x_+)}\label{brb3}\,.
\end{eqnarray}

Eq.~\eq{brb2} coincides with  the equation of motion of a classical test particle of mass $p_\bot$ in the external field $A_+(x_+)$.  In other words,  the test particles effectively move under the action of the longitudinal electric field $E_z=-A_+'(x_+)$. 

\begin{figure}
    \begin{center}
      \includegraphics[width=8cm]{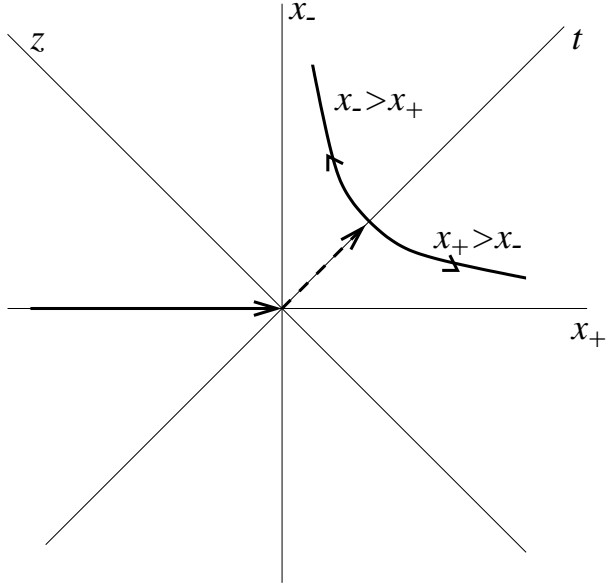}
 \end{center}
\caption{Motion of a particle in the constant background field in the light-cone coordinates. 
Using \eq{brb2} one can derive the trajectory of the particle $x_+x_-=-\frac{p_\bot^2}{2(eE)^2}$ (for simplicity we set $p_+^0=0$).
 At $x_+<0$ the particle moves freely along the light cone $x_-=0$ until the point $x_-=x_+=0$ at which it tunnels 
along the line $x_-=x_+$ (Euclidean path, shown by dashed line) to a real trajectory at $x_+>0$.
 At this point particles move along $x_+>x_-$ branch of the parabola, while antiparticles along the $x_+<x_-$ branch.   
}
\label{part.mot}
\end{figure}

Eq.~\eq{brb3} gives the action of the test particle along the trajectory \eq{brb2}. Its imaginary part arises from the pole in the integrand of the second term in the right-hand-side of \eq{brb3}.
Integration around the pole in the plain of complex $x_+$ yields the imaginary part. It can be calculated replacing the denominator in the first integral   in \eq{brb3} by $\im (p_+^0+gA_+)^{-1}= \pm (i \pi/2) \delta(p_+^0+gA_+)$ according to  the Landau-Cutkosky cutting rule. Additional factor of $1/2 $ arises due to the condition $x_+\ge x_-$. Define 
\beq\label{somedef}
\tau\,=\, x_+\omega\,,\quad A_+(\tau)\,=\,-\frac{E_0}{\omega}f(\tau)\,,\quad 
\gamma\,=\, \frac{p_+^0\omega}{gE_0}\,.
\eeq
where $\omega$ is a typical frequency of the external field and $E(\tau=0)=E_0$. With this definitions we obtain
\beq\label{import}
\im S\,=\, \im\int \frac{p_\bot^2}{gE_0}\frac{d\tau}{\gamma -f(\tau)}\,=\,
\frac{p_\bot^2}{2gE_0}\frac{\pi}{f'(f^{-1}(\gamma))}\,.
\eeq
The imaginary part of the action \eq{import} corresponds to the pair production. In \fig{part.mot} we show a geometrical  interpretation of pair production in the constant background field.

The physical meaning of the \emph{adiabaticity parameter} $\gamma$ introduced in \eq{somedef} is clear: $\gamma=0$ for the static field, 
while $\gamma\gg 1$ for rapidly oscillating one. Since $gE_0\simeq k_{i,+}^2$ and $\omega=k_{i,-}$
 we have the following estimate
\beq\label{nv2}
\gamma\simeq\,\frac{p_+}{k_{i,+}}\,.
\eeq
This estimate for $\gamma$ is the quintessence of a qualitative discussions in Sec.~\ref{sec:2}. Namely, it means 
that for $t=0$ the emission of the gluons is determined by small values of $\gamma$ or,
 in other words, by constant  electric fields, in which $A_+ (x_+)\,= E_0 \,x_+$.

\subsubsection{$\sigma=\pm 1$ case}\label{sec:spinsol}

In the case of $\sigma=\pm 1$ Eq.\eq{sem} cannot be integrated in general.  Equations \eq{ssop1} and \eq{ssop2} remain valid in this case. In place of \eq{ssop2} we obtain
\beq\label{tildep}
\frac{dp_+}{dx_-}\,=\, -gA_+'(x_-)\,-\,\frac{4g\sigma}{p_\bot^2}(p_+-gA_+(x_-))E_z'(x_-)\,,
\eeq
while in place of \eq{ms} we have 
\beq\label{mss}
-2p_-(p_++eA_+(x_-))+p_\bot^2+2g\sigma E_z\,=\,0\,.
\eeq 
Eq.~ \eq{tildep} can be integrated to yield $p_+=p_+(x_-,x_+)$. However, substitution of $p_+$ into \eq{ssop1} gives an ordinary differential equation which cannot be integrated for an arbitrary field $A_+(x_-)$. 

We can still investigate the pair production in the two most important cases of constant and rapidly decreasing fields. 
\begin{enumerate}
\item If $E_z=\mathrm{const}$ then \eq{tildep} reduces to \eq{ssop3}. 
Eq.~\eq{mss} then implies that the solution to
 \eq{sem} with $\sigma=\pm 1$ is given by \eq{import} with  shifted transverse momentum
 $p_\bot^2\rightarrow p_\bot^2+ 2 g \sigma E_z$. This can be seen of course directly in \eq{sem}.

\item 
For large $x_-$ $E_z$ decreases at least as  $1/x_-^{3/2}$ and can be dropped in \eq{tildep} and in \eq{mss} bringing us back to the spinless case \eq{import}.
\end{enumerate}
In both cases sum over spins yields an additional factor of 2 in front of \eq{import}.

\section{Time evolution of the CGC wave function}\label{sec:evol}

\subsection{Model for $A_+(x_-)$}

The $A_+(x_-)$ potential  in \eq{sum2} can in principle be calculated by integrating over the transverse positions $\vec x_{\bot i}$ of the highly energetic partons. 
 However, in the present paper we will restrict ourselves to a simple model which describes both the small and large $x_-$ behavior of the background field.  As was discussed in the introduction, at $t=0$ $E_z=E_0=\mathrm{const}$ which implies that $A_-=-E_0 x_-$.

 There are two important effects determining the late-time behavior of the chromoelectric field. First, the produced gluons start to interact which results in the increase of the gluons' $k_-$ momentum and hence the field frequency $\omega$. We discuss this effect in detail in the following subsection. Second, the produced color pairs screen the original color field. The invariant mass of the pair provides the mass gap in the excitation spectrum. Therefore, we expect the exponential fall-off of the field amplitude.  This can be incorporated in a simple model 
 \beq\label{model}
A_+(x_-)\,=\,\frac{E_0}{\omega}\,(1-e^{-\omega\, x_-})\,.
\eeq
 For this model  we can derive the following expression for the imaginary part of the action, see \eq{import}:
\beq\label{mod2}
\sum_\sigma \im S\,=\, \frac{\pi p_\bot^2}{gE_0(1+\gamma)}\,.
\eeq

The model potential in \eq{model} as well as the simple answer of \eq{mod2} is, of course, a simplification of the real situation. However, it is easy to see that this model incorporates the main properties of the parton cascade that we have discussed.  In \eq{sum2} the density of the color charge can be approximated by 
\beq\label{den1} 
\rho\left(x_-',x_\bot',x_++x_-'-\frac{(x_\bot')^2}{2x_-}\right)\,=\,c\int\,d^2 k_{\perp}e^{i \vec{k}_{\perp} \cdot \vec{x}'_{\perp}}\,\,\delta\left(x'_- -\omega^{-1} \right) \delta\left( k^2_{\perp} \,-Q^2_s \right) \,,
\eeq
where $c$ is a constant. 
Note that before the interaction $x_-$ was negligible since $x_- \sim 1/k_+$. However, right after the interaction its typical value becomes of the order of $x_-\sim \omega^{-1}\sim Q_s^{-1}$ which follows from the uncertainty principle $\Delta k_{+}\,x_{-}\sim 1$ and \eq{nl2} (or \eq{est1}). In writing \eq{den1} we also took into account the fact that most of the gluons have transverse momenta of the order of $k_\bot\sim Q_s$. The density given by \eq{den1} generates $A_{+}(x_-)$ according to \eq{sum2} in the form   
\beq  \label{den3}
 A_{+}(x_{-})\,= c'  \int \,\frac{d^2 \,x'_{\perp}\,d \, x'_{-}}{4\,\pi}\,\frac{2\,x_{-}\,\delta \left(x'_{-} \,-\,\omega^{-1}\right)\,J_0\left(x'_{\perp}\,Q_s \right)}{2x^2_{-}\,\,+\,\,x'^2_{\perp}}   \, =c'\,x_{-}\,K_0\left(x_{-}\,Q_s \right)\,,
   \eeq
where $c'$ is another constant. One can see that \eq{den3} reproduces the main property of the model potential of \eq{model}.  Namely ,
 $ A_{+}(x_{-})\sim x_{-}$ (up to a logarithm) as $x_{-} \,\to\,0$ and $A_{+}(x_{-})\sim\exp\left( - Q_s\,x_{-} \right)$ as $ x_{-}\gg 1$. 
   Therefore, we believe that the model potential  of \eq{model} reflects the main properties of the structure of the parton cascade in high density QCD (CGC). It is worthwhile mentioning that the  mass gap turns out to be of the order of the saturation momentum  and this looks very natural in the CGC approach.

\subsection{Gluon spectrum at $t =0$}\label{sec:init}

To calculate the gluon spectrum we have to calculate the imaginary part of the action $S$ as
 explained in Sec.~\ref{sec:pp}. First, we will calculate the spectrum of produced particles at
 initial time $x_+=0$ and then, in the next section, we will consider later-time particle production. 

It follows from \eq{mod2} that in the constant field ($\gamma=0$)
\beq\label{solqcd}
\sum_\sigma \im S_\sigma\,=\,\frac{\pi p_\bot^2}{gE_0}\,.
\eeq

This equation solves the problem of finding the propagator of a gluon with transverse momentum 
$p_\bot$ in the strong constant classical field.
 In the WKB approach we can guarantee only the exponential suppression for $\im D(Y-y,p_\bot)$ and 
\beq \label{SOL80}
\im D(Y-y,p_\bot) \,\propto\,e^{- 2\sum_{\sigma} \im\mathrm{S}_{\sigma}} \,=
\,\frac{S_P}{\as}\,\ e^{ - \frac{2\pi\,p^2_\bot}{gE_0}}
\eeq
Note that the dependence on the spin $\sigma$ canceled out. Substituting \eq{SOL80} in \eq{PHIPR} we obtain
\beq \label{SOL8}
\varphi(Y-y,p_\bot)\,=
\,S_P\,\frac{4 \pi N_c}{ N_c^2 - 1}\, e^{ - \frac{2\pi\,p^2_\bot}{\,g\,E_0}}\,.
\eeq
The coefficient in front of the exponent in \eq{SOL8}  was chosen based on
the physical meaning of function $\varphi_P$  (see Ref. \cite{NORM}). $S_P$ in \eq{SOL8} is the 
transverse area of
the projectile and $\as$ is the running QCD coupling.

Eq.~\eq{SOL8} allows us to introduce \emph{the saturation scale} which is defined to be the mean momentum
 of the produced gluons:
\beq \label{QS}
Q^2_s \,=\,\frac{g E_0}{2\pi}\,.
\eeq
Using this new variable the unintegrated gluon distribution function becomes 
\beq \label{SOL9}
\varphi_P (p_\bot)\,\,\propto\,\,S_P\,\frac{ \pi N_c}{N_c^2 - 1}\ \,
e^{ - \frac{p^2_\bot}{Q^2_s}}.
\eeq
This equation gives the CGC parton density which coincides with the formula suggested
 by McLerran and Venugopalan in  Ref.~\cite{MV} (see refs.\cite{RA} for more detailed calculation of CGC  parton density), and which illustrates 
the main property of the CGC approach: the entire dependence on rapidity and 
impact parameter enters only through the saturation scale $Q^2_s(y,b)$.

Therefore, our simple picture leads to the CGC initial condition at t=0. In the next section we wish to 
discuss how the system can develop after losing coherence due to the interaction in the final state.

\subsection{Thermalization by a pulse of the chromoelectric field.}\label{sec:pulse}

After losing coherence at $t=0$ the fast gluons start to interact \cite{Kluger:1991ib}. A fast  $i$th gluon in the cascade changes its 
longitudinal momentum and energy according to Newton law
\beq \label{NLFP}
\frac{d k_{iz}}{d t}\,=\,g\,E_z\,\sim\, Q^2_s\,, \quad \frac{d\varepsilon}{dt}= 0
\eeq
The second equation states that the energy of a gluon propagating in the constant background field, which exists at $t=0$, does not change.  Eqs.~\eq{NLFP} imply that during the time of the order of $1/Q_s$ the longitudinal momentum changes its value by $\sim Q_s$. This results in a variation of both $k^{+}_i$ and $k^{-}_i$ by the same value 
\beq\label{nl2}
\Delta k^+_i\simeq\Delta k^-_i \sim Q_s
\eeq
Since $k^{+}_i \gg Q_s$, for $k^+$ it is a small relative change,  and can be neglected. 
However, the initially (at $t=0$) small value of $k^{-}_i= k^2_\bot/k^{+}_i \ll Q_s$ increases in a strong field up to the hard scale $Q_s$.

The 
classical fields will depend on time with the typical frequency of $k^{-}$. Therefore, the interaction among the fast 
partons leads to oscillation of the classical fields with a typical frequency $\omega \,\approx\,Q_s$. However, since 
the values of $k^{+}_i$ for the fast partons are still larger than $Q_s$ we observe that all slow partons (with 
rapidity $y$ in \fig{timecgc} and in \fig{timeclsfield})   stem from the classical emission of the fast partons.

We now turn to the derivation of the gluon spectrum at later times. It was suggested in \cite{KT} that  at later times particles are produced by a pulse of the longitudinal chromoelectric field. Indeed, the third equation in \eq{somedef} implies that the adiabaticity parameter increases with $\omega$. Thus, at later times $\gamma\gg 1$. It follows from \eq{mod2} that in the case of exponentially decreasing field (and \emph{only} in that case) the final spectrum is thermal.
The imaginary part of the action reads
\beq\label{impart2}
\sum_\sigma\im S\nonumber\\
\,= \,\frac{2\pi \, p_-^0 }{\omega}\,.
\eeq
where  we used $p_+^0=p_\bot^2/(2p_-^0)$ which is true for the real particles.

For the imaginary part of the gluon propagator we thus obtain
\beq\label{TER60}
\im D(Y-y,p_\bot)\,=\,\frac{S_P}{\as}e^{-2\sum_\sigma \im S_\sigma}\,=\,\frac{S_P}{\as}\,e^{-\frac{4\pi\,p_-^0}{\omega}}\,,
\eeq
The unintegrated gluon distribution is
\beq\label{TER6}
\varphi(Y-y,p_\bot)\,=\, S_P\frac{4\pi N_c}{N_c^2-1}\,e^{-\frac{4\pi\,p_-^0}{\omega}}\,=\,S_P\frac{4\pi N_c}{N_c^2-1}\,e^{-\frac{p_-^0}{T}}\,.
\eeq
Eq.~\eq{TER6} implies that at later times the gluon spectrum is thermal with the temperature 
\beq\label{temper}
T\,=\,\frac{\omega}{4\pi}\,.
\eeq

\subsection{Thermalization time}\label{sec:therm_time}

The initial state of the heavy ion is characterized by the distribution of gluons \eq{SOL9} with the typical transverse momentum $Q_s$ proportional to the strength of the chromoelectric field, \eq{QS}. Since we have assumed that $Q_s$ is the only relevant scale, the effective temperature $T$ and thermalization time $t_\mathrm{therm}$ (over which the spectrum acquires the thermal shape)  must be related to $Q_s$. To estimate them we will use the following two observations: (i) Due to momentum conservation the $p_+$ momentum gained by the emitted particle is equal to the $p_+$ momentum lost by the field; (ii) The dominant contribution to  the integral of \eq{import} comes from times $\tau\sim\gamma$.  The value of the adiabaticity parameter $\gamma\sim 1$ marks the transition between the gaussian and the thermal distributions.  
In other words,
\beq\label{est1}
\Delta p_+\,=\,p_+(x_-)-p_+^0\,=\, -\omega\,,
\eeq
\beq\label{est2}
\tau\,=\, \omega x_-\,\sim\, 1
\eeq
Eq.~\eq{ssop3}  implies the following estimate  
\beq\label{gain}
\Delta p_+\,\sim\,-\frac{gE_0}{\omega}\,,
\eeq
where in the last equation we used \eq{est2}. Then from \eq{est1} and \eq{gain} we estimate the typical frequency of the field 
\beq\label{freqest}
\omega\,=\,\sqrt{gE_0}\,.
\eeq 
Let us now substitute the definitions of the saturation scale \eq{QS} and the temperature \eq{temper} into \eq{freqest}. The result is
 \beq\label{sattemp}
 T\,\simeq\,\frac{1}{2\sqrt{2\pi}}\, Q_s\,.
 \eeq
The characteristic time over which the field changes is 
\beq
t \,\simeq\,\frac{1}{\omega}\,\simeq\,\frac{1}{\sqrt{2\pi}\,Q_s}\,,
\eeq
and the thermalization time is
\beq\label{thermtime}
t_\mathrm{therm}\,\simeq\,\frac{1}{T} .
\eeq
 In the kinematical range of RHIC for the collisions of heavy nuclei $Q_s \simeq 1 \div 1.5$~GeV \cite{NORM}. This translates to $T \simeq 200 \div 300$~MeV and 
$t_\mathrm{therm} \simeq 0.6 \div 1$~fm.

\section{Nuclear gluon distributions}\label{sec:eA}

\begin{figure}[ht]
    \begin{center}
        \includegraphics[width=0.7\textwidth]{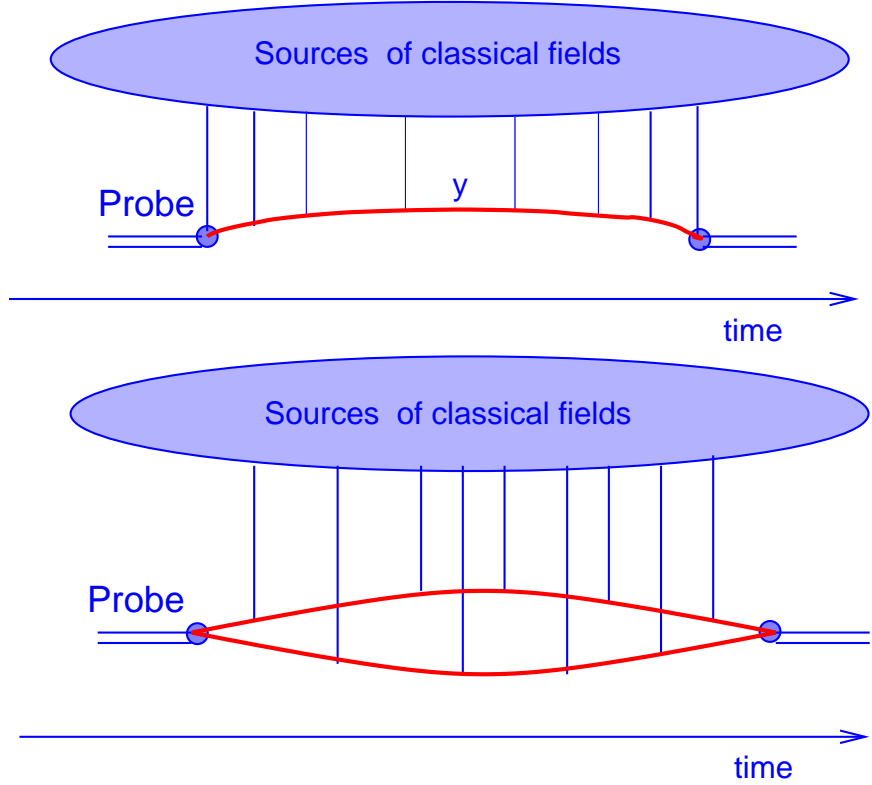}
 \end{center}
\caption{\it The time structure of the deep inelastic scattering
in color glass condensate: the upper figure describes the quasi--classical McLerran-Venugopalan approach while the lower one corresponds to the approach developed in this paper.}
\label{mv}
\end{figure}

To understand better the approximation that we suggest in this paper we consider here the process of deep inelastic scattering off the nuclear target assuming that the nucleus is so heavy that we can 
treat it as a source of the classical field \cite{MV}. Let us assume that the probe is not a virtual photon but is rather a graviton or other particle that can interact with a gluon.
For such a probe we have two different way of interaction with the target. In the first one 
the probe decays into two gluons and one of them  belongs to the classical field of the target (see  
the upper figure in \fig{mv}). The second process goes in two steps: the first is the decay of the probe into two quantum gluons and in the second stage these two gluons interact with the classical field as it is shown in low picture in \fig{mv}.

For the first process we have McLerran-Venugopalan formula \cite{MV}, namely, the distribution of produced gluons in the coordinate space looks as 
\beq \label{MVF}
\frac{d N^{MV}}{d^2x_{\perp}dy}\,\propto\,\frac{1}{\bar \alpha_s}\,\left( 1\,-\, e^{- \frac{1}{4}x^2_{\perp}\,Q^2_s(x)\,\ln (x^2_{\perp}\,Q^2_s(x))} \right)
\eeq
The second process leads to a formula that at first glance  has a quite different form 
namely \cite{MU90}
\beq \label{MVF1}
\frac{d N^{LLA}}{d^2x_{\perp}dy}\,\propto\,\ln(1/x)\,\left( 1\,-\, e^{- \frac{1}{4}x^2_{\perp}\,Q^2_s(x)\,\ln (x^2_{\perp}\,Q^2_s(x))} \right)
\eeq
Eq.~\eq{MVF1} is written in the so-called leading logarithmic $\log(1/x)$ approximation (LLA)  of perturbative QCD
in which we consider only contributions that are proportional to $(\bar \alpha_s \ln(1/x))^n$ such that $\bar \alpha_s\,\ln(1/x) \sim 1$ while $\bar\alpha_s \ll 1$. Since in LLA $\ln(1/x)\sim 1/\bar\alpha_s$ one may conclude that  
\eq{MVF1} gives a contribution of the same order as \eq{MVF}. However, it has been shown in Ref. \cite{LETU} that in the saturation region where $\frac{1}{4}x^2_{\perp}\,Q^2_s(x)\,\geq\,1$ Eq.~\eq{MVF1} 
can be rewritten as follows
\beq \label{MVF2} 
\frac{d N^{LLA}}{d^2 x_{\perp}dy}\,\propto\,\frac{1}{\bar \alpha_s}\ln\left(x^2_{\perp}\,Q^2_s(x)\right)\,\left( 1\,-\, e^{- \frac{1}{4}x^2_{\perp}\,Q^2_s(x)\,\ln (x^2_{\perp}\,Q^2_s(x))} \right)
\eeq
One can see that in the saturation region the contribution of \eq{MVF2} which corresponds to our approach is parametrically  larger than the quasi--classical McLerran-Venugopalan result of \eq{MVF}. It is well known that in the wide region of the kinematic variables the mean field 
approximation to Color Glass Condensate leads to the  geometrical scaling behavior; namely, all experimental obsevables turn out to be functions of one variable $\zeta = \ln\left(x^2_{\perp}\,Q^2_s(x)\right)$, in which we have the non-linear equation. Even
 without discussing the exact form of this equation one can see that \eq{MVF} is the initial condition for such an equation while \eq{MVF2} gives its first iteration. The equation itself \cite{JIMWLK} is  based on the idea that each emitted gluon  with large longitudinal momentum could be treated simultaneously as a quantum
 and as a classical field.  Eq.~\eq{MVF2} is a good illustration of this principle since the quantum emission of  gluons leads to a result with $d N/d^2 x_{\perp}dy \propto \,1/\as$.


\section{Ion-ion collisions}\label{sec:AA}

For ion-ion collisions we intend to use the $k_t$ factorization approach expressed by \eq{GENX}. This equation has not been 
proven for  CGC; nevertheless, we still think that it provides a reasonable starting point, for the following reasons.  First, 
the factorization has been proven for large values of 
transverse momenta \cite{KTFT} (see also reviews in Ref. 
\cite{KTFTST}). Second,  \eq{GENX} is the correct formula for the inclusive production in the case of the BFKL emission 
(see Ref. \cite{LL} and references therein). This fact is very important in understanding why this relation could be valid 
even in the CGC region. Indeed, the BFKL equation has its own, intrinsic scale of hardness: the mean transverse momentum of 
gluons which increases as a function of energy.  This fact is common for the BFKL and CGC emissions, especially if we 
recall that the BFKL approach is the low parton density limit of the CGC. However, the rigorous proof of \eq{GENX} is still lacking. The theoretical situation as well as physical arguments for such factorization have been outlined in Ref. 
\cite{KOVKT} and we cannot add more at the moment.

For the ion-ion collision we thus use the following equation
\beq  \label{IIC1}
\varepsilon \frac{d \sigma}{d^3 p}\, =\,\frac{d \sigma}{d y d^2 p_t}\,=\, \frac{4 \pi N_c}{ N_c^2 - 1}\frac {1}{ p_\bot^2}\,
 \int d k_\bot^2 \,
\alpha_S \, \varphi_{A}(Y - y, k_\bot^2)\, \varphi_{B}(y,
(p-k)_\bot^2),
\eeq
where  $\varphi$ is given by \eq{TER6} and subscripts $A$ and $B$ refer to the mass numbers of the nuclei.
This factorization formula can be rewritten in the form of \eq{FTOUR}, namely,
\begin{eqnarray}\label{IIC2}
\frac{d \sigma}{d y d^2 p_\bot}= \frac{4 \pi N_c}{ N_c^2 - 1} \int d^2 k_\bot &&
\!\!\!\!\im D_A(Y - y,\vec{p}_\bot -\vec{k}_\bot)\,\im D_A(Y -y, \vec{p}_\bot)\,\nonumber\\
&&\!\!\!\!\!\times\, 
\im D_B(y,\vec{p}_\bot - \vec{k}_\bot)\, \im D_B(y, \vec{p}_\bot)\,.
\end{eqnarray}
In the first approximation we can integrate over $k_\bot$ assuming that the classical fields do not depend on the transverse coordinate.
Therefore, we have
\beq \label{IIC11}
\frac{d \sigma}{d y d^2 p_\bot}\,=\, \frac{4 \pi N_c}{ N_c^2 - 1}\, 
\im D_A(Y - y,\vec{p}_\bot)\,\im D_A(Y -y, \vec{p}_\bot)\,
\im\ D_B(y,\vec{p}_\bot)\,\im\,D_B(y, \vec{p}_\bot)\,.
\eeq
 
\begin{figure}[ht]
    \begin{center}
        \includegraphics[width=0.7\textwidth]{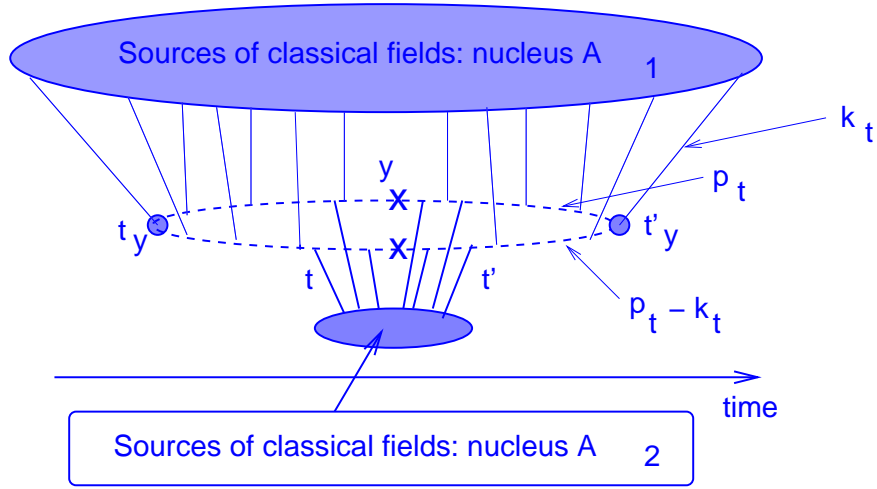}
 \end{center}
\caption{\it Inclusive production in the collision of the nucleus $A$ with the nucleus $B$. Crosses mark the on-mass-shell gluons. 
}
\label{timeA}
\end{figure}

In \fig{timeA} we show that the gluon is moving in the fields $E_A$ and $E_B$ in the time 
interval $t-t'$. In fact, by writing  Eq.~\eq{IIC2} we assumed that the resulting field is just the sum of these two fields. It is correct for QED, but  not  for QCD \cite{MV,JIMWLK,RA}.  In other words, we assumed that during the time interval 
$t - t'$ both gluons interact with two fields in such a way that the resulting propagator is equal to
\beq \label{REPROP}
\im D (t - t') \,=\,\im D_A (t - t')\,\im D_B(t - t')  
\eeq

For $t=0$  \eq{IIC11} leads to
\beq \label{IIC3}                                        
\frac{d \sigma}{d y d^2 p_\bot}\,=\,\frac{S_A\,S_B}{\as}\,\frac{2\,N_c}{N^2_c 
-1}\,e^{ \frac{ -p^2_\bot}{ Q^2_s}}\,;
\eeq
the effective saturation scale in ion-ion collisions thus can be inferred from \eq{IIC3} as $1/Q^2_s=1/Q^2_{s,A}+1/Q^2_{s,B}$
as expected.

For $t > 1/T$ Eq.~\eq{IIC2} looks differently:
\beq \label{IIC4}                                                                                                                                       
\frac{d \sigma}{d y d^2 p_\bot}\,=\,
S_AS_B \frac{4 \pi N_c }{ N_c^2 - 1}\int d^2 k_\bot e^{ - p_-\lb \frac{1}{T_A} +\frac{1}{T_B}\rb}
\eeq
From \eq{IIC4} we see that we have the same expression as in \eq{TER6} but with a different temperature. Therefore the spectrum is given by
\beq \label{IIC5}
\frac{d \sigma}{d y d^2 p_\bot}\,=\,
\frac{S_AS_B}{\as} \frac{ \pi^2 N_c}{ 2( N_c^2 - 1)}e^{-\frac{p_-}{T_\mathrm{eff}}}\,,
\eeq
with
\beq \label{IIC6}
\frac{1}{T_\mathrm{eff}}\,=\,\frac{1}{T_A} \,+\,\frac{1}{T_B}\,.
\eeq

\section{Statistical interpretation of multiple pair production}\label{sec:therm}

\subsection{Probability of multiple pair production}

It was argued in Ref.~\cite{Narozhnyi:1970uv,popov} that the pair production mechanism allows a statistical interpretation. 
Consider the relative probability of single pair production $w_1(\sigma,\vec p)= \exp(-2\,\im S)$. Assuming that the pairs are produced independently, 
the absolute probability to produce one pair is then given by 
\beq
W_1(\sigma,\vec p)\ =\ w_1(\sigma,\vec p) (1 - \sum_{n=1}^\infty w^n_1(\sigma,\vec p));
\eeq
similar expressions hold for the absolute probabilities to produce $n$ pairs, $W_n = w_1^n (1 - \sum_{n=1}^\infty w^n_1)$.
Let $w_0(\sigma,\vec p) = 1 - \sum_{n=1}^\infty w^n_1(\sigma,\vec p)$ be the probability that no pair with quantum numbers $\sigma,\vec p$ is produced. The probability conservation condition then reads
\beq\label{probcon}
w_0(\sigma,\vec p) + \sum_{n=1}^\infty W_n(\sigma,\vec p) = w_0(\sigma,\vec p) \sum_{n=0}^\infty w^n_1(\sigma,\vec p)\,=\,\frac{w_0(\sigma,\vec p)}{1-w_1(\sigma,\vec p)}\,=\,1\,.
\eeq
The total probability that the vacuum of a given theory remains unchanged in a given volume $V$ during time $\Delta t$ is 
\beq\label{vacstab}
W_0\,=\,|\exp(i \mathcal{L}\, V \Delta t)|^2=\exp(-2\,\im \mathcal{L}\, V \Delta t)\,. 
\eeq
On the other hand \cite{Narozhnyi:1970uv,popov}, 
\beq\label{vacstab2}
W_0\,=\,\prod_{\sigma,\vec p}w_0(\sigma,\vec p)\,=\,e^{ \sum_{\sigma,\vec p}\ln (1-w_1(\sigma, \vec p))}\,,
\eeq
where we used \eq{probcon}.
Therefore, 
\beq\label{virial}
\im \mathcal{L}\, V\Delta t \,=\,-\,\frac{1}{2}\,\sum_{\sigma,\vec p}\ln (1-w_1(\sigma, \vec p))\,=\,-\,
\mathfrak{g}\,\frac{V}{2(2\pi)^3}\int d^3p \ln (1-w_1(\sigma, \vec p))\,,
\eeq
where $\mathfrak{g}=(2\sigma+1)(N_c^2-1)$ is the degeneracy factor for pairs of particles. 
The expression on the left hand side of \eq{virial} is nothing but the total production probability in the WKB approximation
\beq\label{prodprob}
1-W_0\,\approx\,
2\,\im \mathcal{L}\, V\Delta t\,.
\eeq
When $w_1$ is given by thermal distribution \eq{TER60} the right hand side Eq.~\eq{virial} is
 related to the  thermodynamic potential $\Omega_\mathrm{pairs}$ of the produced pairs:
\beq\label{sootv}
\Omega_\mathrm{pairs}\,=\,T\,\mathfrak{g}\,\frac{V}{(2\pi)^3}\int d^3p \ln (1-w_1(\sigma, \vec p))\,.
\eeq
Since we work in the approximation in which the background field does not depend on the transverse coordinates the particles produced in a given pair are correlated exactly back-to-back. Therefore, the  
thermodynamic potential  for single particles $\Omega$ is just  twice the one for the pairs.

\subsection{Bose-Einstein condensation}

For the reasons which will become clear shortly, let us introduce a new notation $\mathfrak{W}/T\,=\,2\,\im S$. $\mathfrak{W}$ takes the following values at early and later times
\beq\label{difw}
\mathfrak{W}/T\,=\,\left\{   
\begin{array}{cr}
p_\bot^2/Q_s^2\,, & t\ll t_\mathrm{therm}\\
p_-/T\,,  & t\gg  t_\mathrm{therm}
\end{array}  \right.
\eeq
The number of the produced pairs is equal to \cite{LLV}
\begin{eqnarray}
N\,=\,-\frac{\partial \Omega(\mu)}{\partial \mu}\bigg|_{\mu=0}&=&
\frac{\partial}{\partial \mu} \frac{VT}{(2\pi)^3}\mathfrak{g}\int d^3p \ln(1-e^{(\mu-\mathfrak{W})/T})
\bigg|_{\mu=0}\\ \label{numpair}
&=& \frac{\mathfrak{g}V}{(2\pi)^2}\int dp_\bot^2 dp_z\,\frac{1}{e^{\mathfrak{W}/T}-1}\,,
\end{eqnarray}
where we absorbed the additional degeneracy factor 2 in the definition of $\mathfrak{g}$.

At $t=0$ it follows from \eq{difw}  that the integral  in  \eq{numpair} logarithmically 
diverges in the infrared region in agreement with perturbative QCD. 
However, the total emitted energy is finite
\beq\label{earlyenergy}
\mathfrak{E}\,=\, \int \varepsilon \, dN_\varepsilon\,=\, \frac{V\mathfrak{g}}{(2\pi)^2}\, \Delta p_z \,
 Q_s^3\, gE_0\,
\frac{\sqrt{\pi}}{2}\zeta(3/2)\,,
\eeq
where we have restricted ourselves to the central rapidity region where 
$\varepsilon=p_\bot=\sqrt{2}p_-=\sqrt{2}p_+$, $p_z=0$. We can estimate $\Delta p_z\,\simeq\, gE_0 t$. 
Also during the longitudinal expansion $V\sim t$. Therefore, during the early stages after the collision 
the energy flows from the field to the soft particles as $\mathfrak{E}\sim t^2$.  

The gluon number becomes finite as soon as $t>0$. Indeed, when $t$ changes from  $0$ to $t_\mathrm{therm}$  it follows from \eq{difw} that $n=d\ln\mathfrak{W}/d\ln p_\bot$ decreases from $2$ to $1$. We have
\beq\label{internumber}
N\,=\,\frac{2V\mathfrak{g}}{(2\pi)^2}\int dp_\bot  dp_z\,\frac{p_\bot}{e^{(p_\bot/\Lambda)^n}-1}\,=\,
\frac{V\mathfrak{g}}{(2\pi)^2}\,gE\,t\,\Lambda^2\,F(n)\,.
\eeq 
where $\Lambda$ varies from $Q_s$ at $t=0$ to $T$ at $t=t_\mathrm{therm}$  and
\beq\label{di}
F(n)\,=\,\int_0^\infty \frac{dz\, z}{e^{z^n}-1}\,,\quad F(1)=\frac{\pi^2}{6}\,.
\eeq
The distribution in \eq{di} at $n=1$ has  the form of a Bose-Einstein distribution with a vanishing 
chemical potential,  $\mu=0$.  We thus expect the Bose-Einstein condensation of gluons to occur at temperatures lower than the critical temperature $T_0$. For the sake of simplicity let us assume that $t$ is close to $t_\mathrm{therm}$ so that $n\gtrsim 1$ and $\Lambda\gtrsim T$. To calculate $T_0$ let us note that \eq{internumber} cannot be used for counting the number of particles which carry zero transverse momentum $p_\bot=0$ at $T<T_0$ \cite{LLV}, where $T_0$  is defined as  
\beq\label{condtemp}
T_0\,=\,\left( \frac{3(2\pi)^2 N}{V\mathfrak{g}\, gE\, t\, \pi^2} \right)^{1/2}\,.
\eeq
 The number of particles with zero momentum (in the condensate) equals
\beq\label{incond}
N(p_\bot=0)\,=\,N\,\left[ 1- \left(\frac{T}{T_0}\right)^2\right]\,,
\eeq
whence $N$ is the total (finite) number of particles. 

The critical temperature  decreases with time.   Let us estimate its value at $t=t_\mathrm{therm}$.  Using \eq{QS} and \eq{thermtime} we get $gE_0t_\mathrm{therm}\sim 1$. Assume $V=S_A t$ where $S_A$ is the transverse cross sectional area of the nucleus (for simplicity we assume a
 central collision). Then 
\beq\label{tmax}
T_0\,\le\,T_0(t_\mathrm{therm})\,=\,
\,\left( 
\frac{12N}{\mathfrak{g}\, S } 
\right)^{1/2} \,.
\eeq
The total number of hadrons produced at $y=0$ at RHIC is about $N\sim1000$. Using $S_A=\pi (7~ \mathrm{fm})^2$ 
and $\mathfrak{g}=2\cdot3\cdot 8$ we obtain $T_0(t_\mathrm{therm})\approx 260$~MeV. 
 Therefore, just after the system is thermalized, a significant fraction of gluons may form a Bose-Einstein condensate.

The Bose-Einstein condensation of soft gluons in high energy QCD leads to a remarkable consequence. 
Recall that  the typical correlation length  inside the high energy hadron is rather small $\lambda_\mathrm{c}\simeq 1/Q_s$. This implies that the gluon emission with long wavelengths $\lambda=1/p_\bot>\lambda_\mathrm{c}$ is suppressed because 
it decouples from the hadron wave function, similarly to the decoupling of a large wavelength signal from a small antenna. 
Therefore, one is led to predict a  deficit of soft gluons at high energies, in a stark contradiction with the experimental data. The phenomenon of Bose-Einstein condensation solves this puzzle since it allows piling up of soft gluons. 
 
\subsection{Viscosity of the parton system}

We have argued in Sec.~\ref{sec:pulse} and Sec.~\ref{sec:therm_time} that at $t>t_\mathrm{therm}$ the produced partons have 3D thermal distributions with an effective temperature $T$.  The number of produced particles per unit volume is large $n\sim 1/\as$ since they were part of the classical fields in the initial wave functions before the collision. This observation is an important argument in support of the hydrodynamical description of  the parton system at later 
times \cite{Hirano:2004rs}. 

The typical transverse momentum of a parton is $\langle p_\bot \rangle \sim T$.  Recall that the temperature $T$ is proportional to  the saturation scale $Q_s(y)$ which is an exponential function of rapidity. Therefore, the temperature varies with rapidity. As a consequence, the average value of transverse momentum $\langle p_\bot \rangle $ significantly varies between different  rapidity layers. The 
difference in the transverse momentum distributions along the longitudinal axis of rapidity amounts to the  
shear viscosity\footnote{We would like to thank Ben Svetitsky for bringing our attention to this consequence of our approach.}.

The shear viscosity can be estimated as (we keep only parametric dependence while omitting all numerical factors)
\beq\label{vis}
\frac{\eta}{n}\,=\, \langle p_\bot \rangle \, \lambda\,\sim\, \frac{Q_s}{n\,\sigma} \,,
\eeq
where $\sigma\sim \as/Q_s^2$ is the scattering cross section for a parton in the classical background field. 
The number of particles per unit volume is 
\beq\label{conc}
n\,\sim\, \frac{xG}{S_A L_z}\,,
\eeq
where $L_z\sim 1/Q_s$ is the longitudinal extent of the system. Using $Q_s^2\sim \as xG/S_A$ we then estimate
\beq\label{estv}
\frac{\eta}{n}\,\sim\,1\,.
\eeq
This estimate implies the parametric smallness of viscosity which comes about as a consequence of high occupation number of gluons in the initial wave function. In contrast, in pQCD the shear viscosity is parametrically enhanced $\eta/n \sim 1/\as^2$ \cite{Arnold:2000dr}.

\section{Discussion and summary of the results}\label{sec:discussion}

In this paper we have developed an approach to particle production based on the principle idea of CGC: the gluon with a
 rapidity $y_0$ can be considered as emitted 
by the classical fields that are composed of all faster partons with $y > y_0$. We showed that in such an approach 
the gluons at the moment of collision are 
emitted by the classical longitudinal fields ($E_z$),  which are created by fast particles during 
very short time $\sim 1/\varepsilon$ after the collision ($\varepsilon$ is the particle energy in
the laboratory frame).
We found the relation \eq{QS} between the momentum scale of dense partonic system $Q_s$ and
the strength of the  classical field $E_z$. 
The inclusive distribution  at $t =0$ is given by \eq{SOL9} and  turns out to be the same as has been 
expected in the CGC approach (McLerran-Venugopalan model) \cite{MV}.

At later times $ t \gtrsim 1/\sqrt{2 \pi}\,Q_s$, one has to consider the time dependence of the classical fields. 
We followed through the evolution of the system assuming that the main source of the produced gluons is still the classical 
field created by faster partons. 
 It turns out that the momentum distribution of the produced gluons has  a three--dimensional   
thermal spectrum given by \eq{TER6} with $T = (1/4\sqrt{2 \pi}) Q_s(y)$ 
for the collision of two identical nuclei at midrapidity.
Therefore, the CGC approach led to a thermal spectrum of emitted gluons with an effective  temperature 
which depends on the rapidity of emitted 
gluons. 

It was argued in Ref.~\cite{KOV} that the perturbative dynamics may not be adequate for the description of the late-time processes in a high-energy heavy-ion collisions as it does not lead to the thermalization as anticipated on general grounds. 
In the present paper we circumvent that result by suggesting a non-perturbative mechanism of thermalization. The non-perturbative nature of the obtained results can be clearly seen in Eq.~\eq{import} which exhibits non-analytic dependence on the coupling $g$.

The dependence of  temperature on rapidity may trigger instability of the gluon
 system (see for example \cite{Arnold:2004ti,Arnold:2004ti,Romatschke:2005pm} and references therein) and speed up thermalization process. Perhaps at late times the instability
 driven thermalization can compete with pair production by strong fields discussed in this paper.
 This problem warrants further investigation. 

Another problem left beyond the scope of the present paper is understanding at what time 
the hydrodynamic description becomes valid. It seems reasonable to asssume that for times later  
than $t_\mathrm{therm} \gtrsim 1/T$ we could apply the viscous  hydrodynamic description. Indeed, we showed that for these times we have a 3D thermal distributions in each 
slice of 
rapidity which is a pre-condition for using the hydrodynamic approach. On the other hand, the average 
transverse momenta $\langle p_t\rangle \,\simeq\,T$ are quite different in the two neighboring slices in rapidity 
due to the dependence of $T$ on rapidity. Therefore, we can expect a considerable 
difference in parton momentum distributions in different rapidity slices
 which amounts to viscosity.  We have argued that the CGC initial conditions lead 
to the parametrically small shear viscosity $\eta \sim {\mathcal O}(1)$ as opposed to the perturbative result, $\eta \sim {\mathcal O}(1/\as^2)$. 
 It should be mentioned that
 matching the CGC energy-momentum tensor with that of an almost
 perfect fluid yielded similar results \cite{Kovchegov:2005az}.
 
Although we performed our calculations for the $SU(2)$ gauge theory, we believe that all the qualitative features of the  derived results will remain valid for the realistic color group $SU(3)$ as well. Calculations of the pair production effect in a constant chromoelectric field of $SU(3)$ have been recently done in Ref.~\cite{NN,Gelis:2005pb}. Unlike $SU(2)$  there are two Casimir operators in  $SU(3)$ which yield a more complicated dependence of the pair production effect on $E$.

A new related general approach to particle production in field theories coupled to strong external sources  has been recently formulated in Ref.~\cite{Gelis:2006yv} where the particular example of $\lambda\phi^3$ theory has been discussed. It may yield new insights into the problem of particle production problem in QCD as well.

It is interesting to note that calculation of inclusive $e^+e^-$ production in QED can be done in 
exactly the same way as was followed to calculate the gluon production in this paper. Indeed, a fast moving system
 in QED is characterized by large transverse fields which lead to 
bremsstrahlung production of photons which is a classical process.  There is also  production of
 $e^+e^-$ pairs which is a typical quantum process. The QED variant of the CGC approach states that at high energies the 
  inclusive production is dominated by the emission of  $e^+e^-$ pairs in the classical photon field and not by the 
  quantum emission of virtual photons.

\acknowledgments
We want to thank Gerald  Dunne, Asher Gotsman, Yuri Kovchegov, Tuomas Lappi, Larry McLerran, Gouranda Nayak, Uri Maor, Jianwei Qiu, James Vary and Raju Venugopalan for useful discussions on the subject of this paper. The research of D.K. was supported by the U.S. Department of
Energy under Grant No. DE-AC02-98CH10886.
K.T. would like to thank RIKEN, BNL and the U.S. Department of Energy (Contract No. DE-AC02-98CH10886)
 for providing the facilities essential for the completion of this work.
This research was supported in part  by the Israel Science Foundation,
founded by the Israeli Academy of Science and Humanities and by BSF grant \# 20004019.

\appendix{
\begin{boldmath}
\section{Equation of motion of a vector particle in an external field in $SU(2)$}
\end{boldmath}
\label{appendix}

Let $A_\mu=A_\mu^3$ be the background classical field. We are looking for the equations of motion of the vector particle $W_\mu=(A_\mu^1\,+\,iA_\mu^2)/\sqrt{2}$ in the background field $A_\mu$. The 
Lagrangian is
\beq\label{lagrsu2}
\mathcal{L}\,=\,-\frac{1}{4}\sum_{i=1}^3 \left(F_{\mu\nu}^i\right)^2\,. 
\eeq
Using identity
\begin{eqnarray}\label{lagr3}
&&\sum_{i=1,2}(F_{\mu\nu}^i)^2\,=\,2\,|D_\mu\,W_\nu\,-\,D_\nu\,W_\mu|^2\,=\,\nonumber\\
&& 
\big[(\partial_\mu A_\nu^1-\partial_\nu A_\mu^1)+g
(A_\mu^3A_\nu^2-A_\nu^3 A_\mu^2)\big]^2+
\big[(\partial_\mu A_\nu^2-\partial_\nu A_\mu^2)-
g(A_\mu^3 A_\nu^1-A_\nu^3 A_\mu^1)\big]^2\,,
\end{eqnarray}
and expanding the 3-component of the strength tensor 
\beq
(F^3_{\mu\nu})^2\,=\,(\partial_\mu A_\nu^3-\partial_\nu A_\mu^3)^2\,+\,
2g(\partial_\mu A_\nu^3 - \partial_\nu A_\mu^3) \, 2  i\,  W_\mu^*\,W_\nu \,+\,
\mathcal {O}(W_\mu^4)
\eeq
we obtain 
\beq\label{lagr.su2}
\mathcal{L}\,=\, -\frac{1}{4}\,(\partial_\mu\,A_\nu\,-\,\partial_\nu\,A_\mu)^2\,-\,i\,g\,
(\partial_\mu\,A_\nu\,-\,\partial_\nu\,A_\mu)\,W_\mu^*\,W_\nu\,-\frac{1}{2}\,|D_\mu\,W_\nu\,-\,D_\nu\,W_\mu|^2\,,
\eeq
where $D_\mu=\partial_\mu+igA_\mu$.  
The corresponding equation of motion is
\beq\label{eq.m.su2}
[-\,D_\lambda^2\,\delta_{\mu\nu}\,+\,D_\mu\,D_\nu\,+\,2\,i\,g\,F_{\mu\nu}]\,W^\nu\,=\,0\,.
\eeq

Assuming $D^\mu W_\mu=0$ we have
\beq\label{f.ex}
\partial_\mu^2\,W_\nu\,-\,2\,i\,g\,A_\mu\,\partial^\mu\,W_\nu\,-\,i\,g\,(\partial^\mu\, A_\mu)\, W_\nu\,-\, g^2\, A_\mu^2\, 
W_\nu\,+\, 2\, g\, E\, \sigma\, W_\nu\,=\,0\,,
\eeq
which is equivalent to \eq{EQM}.}


\end{document}